# THE INSPIRING UNIVERSE

**George Miley[1,3], Carolina Ödman[2,4] and Pedro Russo[1,5]**

[1] Sterrewacht Leiden, Leiden University, The Netherlands
[2] carolina.odman@gmail.com
[3] Initiator and International Coordinator
[4] International Project Manager 2005 – 2010
[5] International Project Manager 2011 – 2017

# 1. INTRODUCTION

## 1.1 Preamble

Humans like to find things out. They naturally ask questions and explore the world around them. One observes such behavior already in neonatal infants. Inspiration to further develop such exploratory behavior, the quest to understand and do so deeply, is stimulated by different things, depending on the environment in which one is born and grows up. We thus do not all have the same chances to be inspired. However, there is one exception. Throughout human history humans around the world have stood in awe of the wonders of the sky, the universe. This is a source of inspiration that unites us all. The above considerations and possibly other ones have led to the Universe Awareness (UNAWE) project, which focuses specifically on underprivileged children between the ages of four and ten. Universe Awareness (UNAWE) uses the beauty and grandeur of the Universe to encourage young children, particularly those from an underprivileged background, to have an interest in science and technology and foster their sense of global citizenship from the earliest age.

Key precepts of the *Building the Scientific Mind* approach to education are (i) to cultivate habits of rational thought and (ii) to stimulate the acquisition of ethical values. These principles are also at the heart of the Universe Awareness philosophy. In this chapter we shall outline the rationale on which UNAWE is based and describe the various elements and accomplishments of UNAWE.

The famous adage "Give me a child until he is 7 and I will give you the man" is attributed to Francis Xavier, 16th century co-founder of the Jesuit Order. Except for the implicit sexism of the time, this quotation is still very relevant for education. To influence the developments of a person's character and values, it is important to begin at an early age. As argued e.g. by Gopnik, Meltzoff, & Kuhl (1999), babies and young children are curious and this natural curiosity should be stimulated. By the age of 7, many seeds of the basic attitudes and values that will govern the behavior of the future adult will have been sown. The nature of these values can determine whether a society is tolerant or intolerant, exclusive or inclusive. It is not coincidental that many world religions regard the control of schools for young children as crucial to their existence. The philosophy that governs the education of young children can strongly influence the value systems and ethical standards of society.

## 1.2. Astronomy and space as educational tools

Astronomy is an effective holistic educational tool because it embodies a unique combination of science, technology and culture (Figure 1). The cultural roots of astronomy satisfy the deepest philosophical yearnings by observing far into our past universal history. Unraveling the history of the Universe has been one of the great human achievements of the last half-century.

In the 21st century, astronomy has become a gateway to physics, chemistry, biology and mathematics and a driver of advanced developments in electronics, optics and information technology. Ground-based and space observatories have opened up a golden age of astronomy. Many profound discoveries have been made that has changed our view of the Universe. We have learned that some objects that inhabit our Universe are so strange that their existence could not have been predicted. Others are so spectacular that they stretch the boundaries of human wonder. The beautiful and intriguing images produced by high-tech telescopes and satellites can capture the imagination of children of all ages. Astronomers continue to push the frontiers of knowledge with global scientific collaborations. During the past decade the discovery of planets outside our Solar System and the realization that many billions of planets must inhabit our Galaxy has profound implications for human civilization.

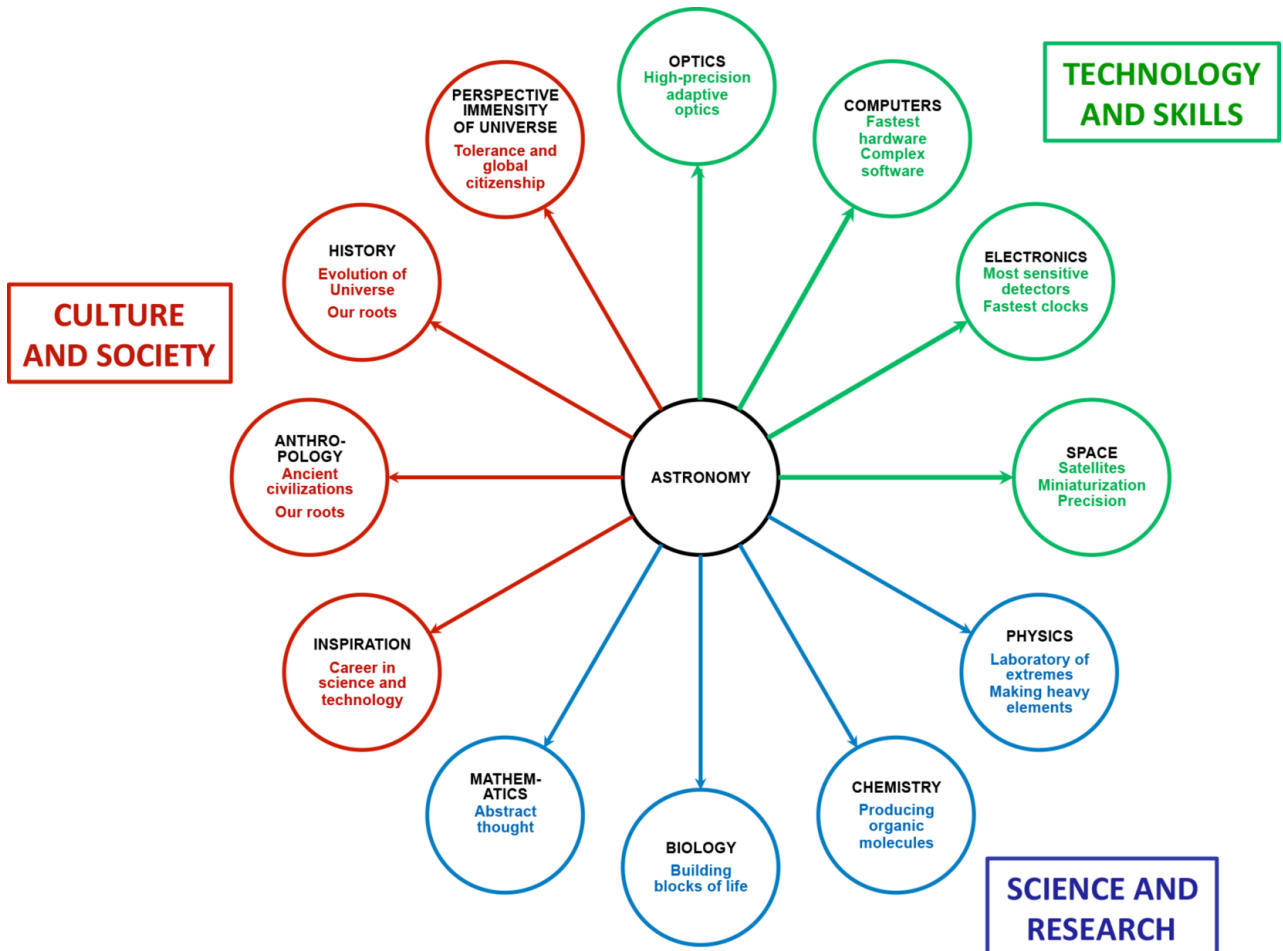

**Figure 1:** *The wide applicability of astronomy and space sciences as an educational tool and a gateway to science, technology and culture (International Astronomical Union Strategic Plan: 2011, Miley 2012)*

**2. THE UNIVERSE AWARENESS PROGRAMME**

**2.1 Astronomy as a motivator of young children.**

The excitement of astronomy and the adventure of space are particularly important vehicles for capturing the imagination of young children, introducing them to science and engineering and helping to stimulate their ethical value systems. There are several reasons for this.

- Consideration of the vastness and beauty of the Universe and of our place within it provides a special perspective that can help broaden the mind and stimulate a sense of global citizenship.
- The ability of space and astronomy to interest and *motivate* young children can be used to stimulate education in a broad holistic sense. Space-based themes can be encapsulated in stories to further language skills and to pose and solve problems that develop numerical skills.
- Astronomy provides a seductive introduction to science and technology. Many scientists and engineers trace their first interest in science to exposure as very young children to the fascinating Universe. Although the main goals of Universe Awareness are not directed at producing future scientists, exposure to intriguing aspects of space can 'light a spark' in a young child that several years later causes her to embark on a scientific or technical career.

**2.2 Genesis of Universe Awareness**

Based on the rationale presented above, UNAWE was initiated in 2005 to exploit scientific, educational and social dimensions of astronomy for the education of young children (Miley, Madsen, & Scorza, 2005). Directed at children between 4 and 10 years, UNAWE aims to broaden children's minds, awaken their imagination and curiosity in science and encourage respect, tolerance and global citizenship. A major goal is to stimulate children to develop into curious, tolerant and internationally oriented adults (Ödman, Scorza, Miley, & Madsen, 2006; Ödman, 2007; 2011).

The main ingredients of UNAWE are:
- Provision of an international network for astronomy as an educational tool for motivating very young children.
- Development of country-specific educational resources,
- Organization and stimulation of training for teachers,
- Development and implementation of a consistent evaluation framework.

UNAWE is based on the philosophy that the early formative years are crucial in the development of the human value system. From the age of 4 children can readily appreciate and enjoy the beauty of astronomical objects and can learn to develop a 'feeling' for the vastness of the Universe. On the one hand, the discovery of the universe is an excellent and exciting introduction to the scientific method and the concept that nature can be interrogated by rational means. On the other hand exposure to inspirational astronomical observations and space data can help broaden the minds and stimulate a world-view. The enormity of the Universe is also a subject for deep wonder. In particular, the Universe contains countless exotic objects that are ideal for feeding a child's imagination. To quote the European biologist J. B. S. Haldane, "The Universe is not only queerer than we suppose, but queerer than we can suppose" (Haldane, 1927, p. 286). The strange conditions that exist elsewhere in the Universe and beautiful pictures from modern telescopes are an ideal backdrop for UNAWE activities.

Studies of other planets in our solar system provide important lessons relevant to protecting the environment of the earth. To quote Carl Sagan (1980 [1990 update]): *"The discovery that such a thing as nuclear winter was really possible evolved out of the studies of Martian dust storms. The surface of Mars, fried by ultraviolet light, is also a reminder of why it's important to keep our ozone layer intact. The runaway greenhouse effect on Venus is a valuable reminder that we must take the increasing greenhouse effect on Earth seriously."*

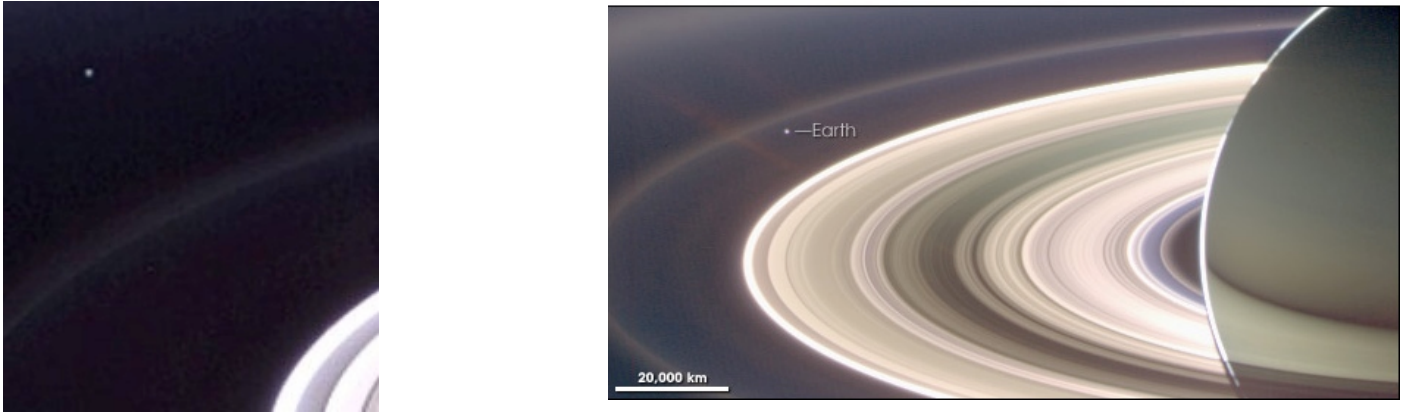

**Figure 2:** *An iconic picture of the Earth as Sagan's "faint blue speck" (blow-up on left). The ESA-NASA Cassini spacecraft took this photo from a location close to Saturn, whose rings are shown. The blow-up is part of the larger Cassini picture of Saturn (right). Courtesy NASA.*

**2.3. Why does UNAWE emphasize underprivileged communities**?

Although the resources of UNAWE are open to all, UNAWE concentrates on children from relatively underprivileged communities. The rationale for this is threefold:

1. Early educational interventions during the young formative years are crucial in stimulating child development (OECD, 2006; Worth & Grollman, 2003; Rocard, 2007).
2. There is an enormous untapped potential and talent in underprivileged communities. Such communities are frequently alienated from society at large and this alienation begins at childhood. According to a UNICEF report (UNICEF, 2006), *"The opportunity to help disadvantaged children have a more equal start in schooling is in the earliest years when the basis for their cognitive, social and emotional development is being formed"* (p. 1).
3. Educating very young children from underprivileged communities is highly cost effective (Schweinhart, Mont, Xiang, Barnett, Belfield, & Nores, 2005), but needs a special approach (Early Years Learning Framework for Australia, Commonwealth of Australia, 2009; Arnold and Doctoroff: 2003).
   A series of important studies by James Heckman, recipient of the Nobel Memorial Prize in Economic Sciences, and his group emphasizes the cost effectiveness of early educational intervention in underprivileged communities with pre-school activities activities that motivate very young children (Cunha, Heckman, Lochner, & Masterov, 2006; Heckman & Masterov, 2007; and Heckman, 2000; 2008). See Figure 2. However, it is clear that further such studies are needed in a range of diverse environments and cultural settings.

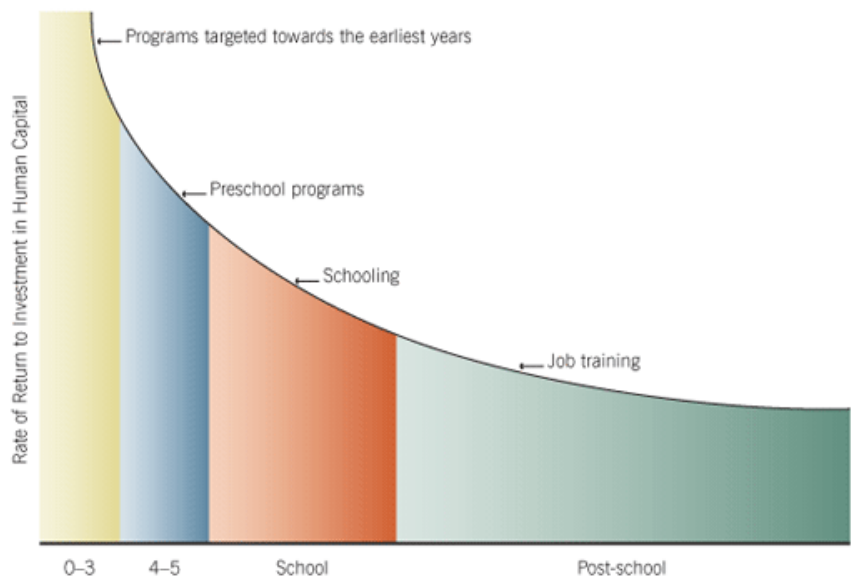

**Figure 3:** *This graph demonstrates that rates of return on human capital investment decreases with age, with the highest return on investments at preschool age. (Heckman, James J. "Schools, Skills and Synapses," Economic Inquiry, 46(3): 289-324 (2008).*

**2.4. Why does UNAWE target teachers?**

One of the most important components of UNAWE programs is teacher training. There are several reasons for this.
- The first is to increase the reach of the program. The effectiveness of reaching children is multiplied by a large factor through targeting their teachers. Enthusiastic teachers can 'infect' thousands of children with their enthusiasm.
- The second reason is that, while elementary school teachers have a substantial influence on the first school years of young children, they often feel inadequate to teach STEM (science, technology, engineering and mathematics) issues (e.g. Peacock, 1991; Beilock, Gunderson, Ramirez, & Levine, 2009). Universe Awareness training is designed to build the confidence of teachers in teaching these topics. We demonstrate to teachers that space can be an important tool for motivating young children and introducing them to the excitement of science and technology. After they are motivated, this excitement can also be exploited e.g. to teach children language skills through reading space stories or math skills through solar system problems.
- An indirect factor in reaching out to teachers is that elementary school teachers are highly undervalued in many countries. Membership of the international (multidisciplinary) UNAWE network can help enhance the prestige of school teachers, particularly in some developing countries.

**2.5. Furthering gender equality.**

In many parts of the world, women show less interest in science and engineering than men. Several reports and studies have identified reasons why girls are significantly more negative about certain science subjects and careers than boys and traced the views and attitudes of boys and girls about science to influence from an early age (e.g. Osborne & Collins, 2000). Gender imbalance needs to be addressed at all levels of education and all career stages. It is crucial to dispel the myth that girls can't do science and that technical subjects are better left to the boys.

Since its inception UNAWE has had furthering gender equality as an important goal. The education and upbringing of very young children can play a substantial role in helping to tackle gender imbalance. During ages from 4 to 10 years the value systems of children form and preconceptions about gender roles begin to take shape. Furthermore, in most countries throughout the world elementary school teachers tend to be women. By building up the confidence of these teachers in science-related topics, the proposed teacher training courses will contribute to projecting an image of women who are confident in science and engineering.

### 3. GROWTH AND ACCOMPLISHMENTS OF UNAWE

### 3.1 UNAWE - a global activity.

Universe Awareness was developed as an international partnership with a bottom-up structure. The program is unique in that it is the only global project that links top international researchers to the education of young children and the training of their teachers. Local experts drive the implementation of the program in each partner country and give special attention to local cultural needs and histories. Since its initiation in 2005 UNAWE has grown into a multidisciplinary network with more than 500 participants and is active in more than 57 countries worldwide (see Figure 4), with more than 22 different languages.

UNAWE was adopted as a cornerstone program of the UN-ratified International Year of Astronomy 2009 and an official activity of the International Astronomical Union. At its inception UNAWE was endorsed by several Nobel Prize winners (for example Prof. Riccardo Giacconi, 2002 Nobel Prize winner; Prof. Joseph Taylor: 1993 Nobel Prize winner, Prof. Charles Townes: 1964 Nobel Prize winner). UNAWE has also received considerable attention internationally, including a feature article in Physics Today, the professional magazine of the American Institute of Physics (Feder, 2007). A highlight was the award to UNAWE of the 2011 Science Magazine Science Prize for Online Resources in Education (SPORE) (Ödman-Govender & Kelleghan, 2011). Because of its success, UNAWE was mandated as expert for e.g. the UNESCO World Report: "*Investing in Cultural Diversity and Intercultural Dialogue*" and invited to participate in the setting up of an Outer Space Committee of the Conference of NGOs in consultative relationship with the United Nations (CONGO).

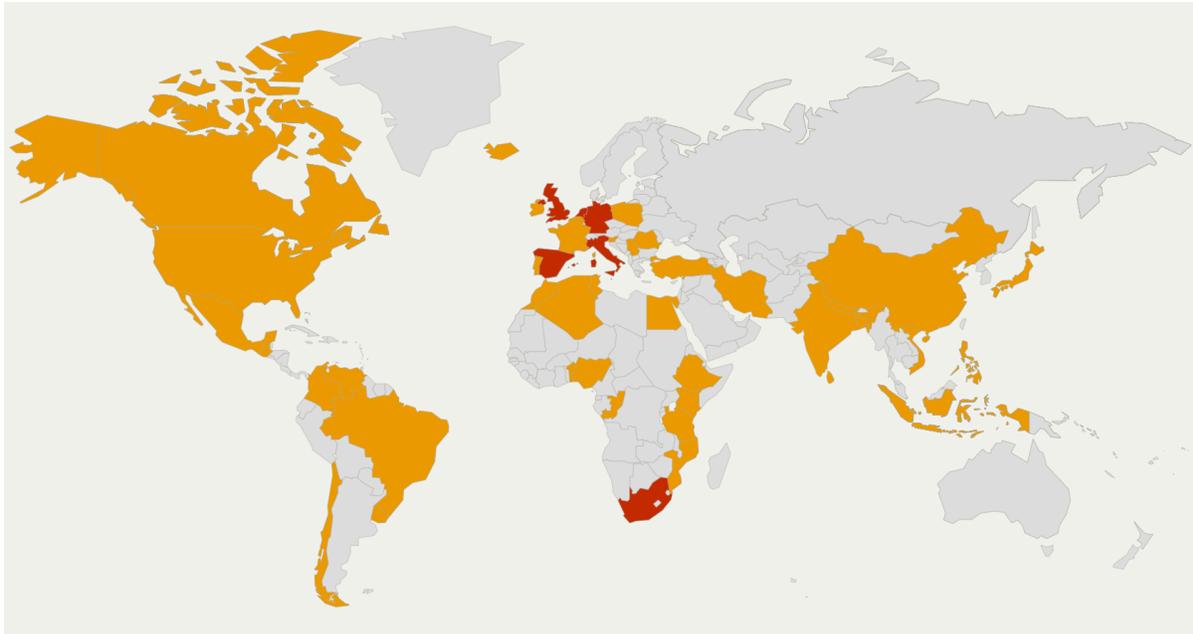

*Figure 4:* *The 57 UNAWE member countries (in brown), and the 6 countries participating in the FP7 DG Enterprise and Industry EU-UNAWE project in red.*

In 2010, UNAWE received a considerable boost with the award of a 2 million Euro 3-year project by the space program of the European Commission. "EU Universe Awareness (EU-UNAWE) is implementing substantial UNAWE programs in five European countries (Germany, Italy, the Netherlands, Spain, and the UK) and South Africa (e.g. Russo, Kimble et al. 2011; Schrier, Nijman, & Russo, 2012). EU-UNAWE exceeded its original goals by a wide margin. EU-UNAWE trained almost 1100 teachers and reached more than 60,000 pupils through direct and indirect activities and was evaluated highly by independent reviewers appointed by the European Commission.

During the course of EU-UNAWE extensive an assessment was conducted among teachers and educators in all six participating countries to study the effectiveness of the teacher training and the resources. It was concluded that the project was highly effective and efficient in developing motivation, critical thinking, knowledge and interest in space, and intercultural attitudes.

**3.2 Universe Awareness resources and their development**

The development of suitable resources for teachers is an essential element of UNAWE. These resources attempt to assist teachers to deal with important concepts while at the same time emphasizing the wonder and excitement of the Universe.

A major achievement of the EU-UNAWE project has been to create a repository of open-source age-related educational materials in several languages for teaching children ages 4 to 10 and to make them available freely to teachers and parents via various national websites as well as the UNAWE international website. UNAWE resources undergo stringent peer review to ensure their pedagogical quality and accurate scientific content. This review process developed in the context of the EU-UNAWE project was adopted by the "Task Force for Schools and Children" of the International Astronomical Union Office of Astronomy for Development and is embodied in the IAU-endorsed *astroEDU* platform for educational resources (http://www.iau.org/astroEDU) (Russo, Miley, Kimble et al., 2015).

Almost 200 resources have been produced to date. UNAWE 'flagship' resources are the popular *UNAWE Earthball* and *Universe in a Box*. The *UNAWE Earthball* is a simple inflatable globe without borders. It shows the earth from the perspective of Space and illustrates important aspects of our planet such as its shape, its axial tilt and its surface composition. The accompanying activity booklet contains 50 interactive, hands-on activities for children aged between 6 and 12 years that can be put into practice in schools, playgrounds or at home. They cover a broad range of topics, including astronomy, geography, environmental awareness, global citizenship and even biology.

*Universe in a Box* is an educational kit that provides teachers and educators with more than 40 practical activities as well as the materials and models required to carry them out. It has a modular design with five modules: Our Fascinating Moon; The Earth — Our Home Planet; The Sun — Our Home Star; Our Solar System; and The World of Constellations. In accordance with the UNAWE philosophy these were designed to

- Encourage inquiry-based learning among children, by involving observing, discussing, drawing conclusions and presenting.
- Link astronomical topics with other subjects such as mathematics, art, and philosophy to support interdisciplinary learning and present a more holistic view of our universe.
- Make children aware that they are inhabitants of a small fragile planet.
- Foster respect for other cultures.

Another accomplishment of EU-UNAWE was the development of 'Space Scoop.' Space Scoop is an astronomy news service for children aged 8 and above. The latest press releases from leading astronomical ground and space observatories are rewritten, so that they are understandable for children (Russo et al. 2013 -http://www.unawe.org/resources/spacescoop/). The Space Scoop case study is described in more detail in the next section.

**4. UNIVERSE AWARENESS CASE STUDIES IN INNOVATION**

**4.1 Case Study 1: Teacher Training – Duo Internships**

During 2009 -2010 and in 2014, the Universal Awareness project organized and coordinated a pilot project in teacher training in The Netherlands involving 'duo-internships.' During these internships, pairs of student primary school teachers and astronomy university students were coupled and charged with developing astronomy lessons and implementing these lessons at primary schools. Nineteen such 'duos' participated in activities at primary schools in Groningen, Arnhem, The Hague, Rotterdam and Utrecht. The combination of an astronomy student and a student teacher ensured that lessons were both scientifically correct and pedagogically sound. Moreover, the students learned from each other. The astronomy students learned how to explain science in a way that the children could understand and the aspiring teachers learned to master some simple inspirational scientific concepts. These duo-internships were experienced by all concerned as highly successful.

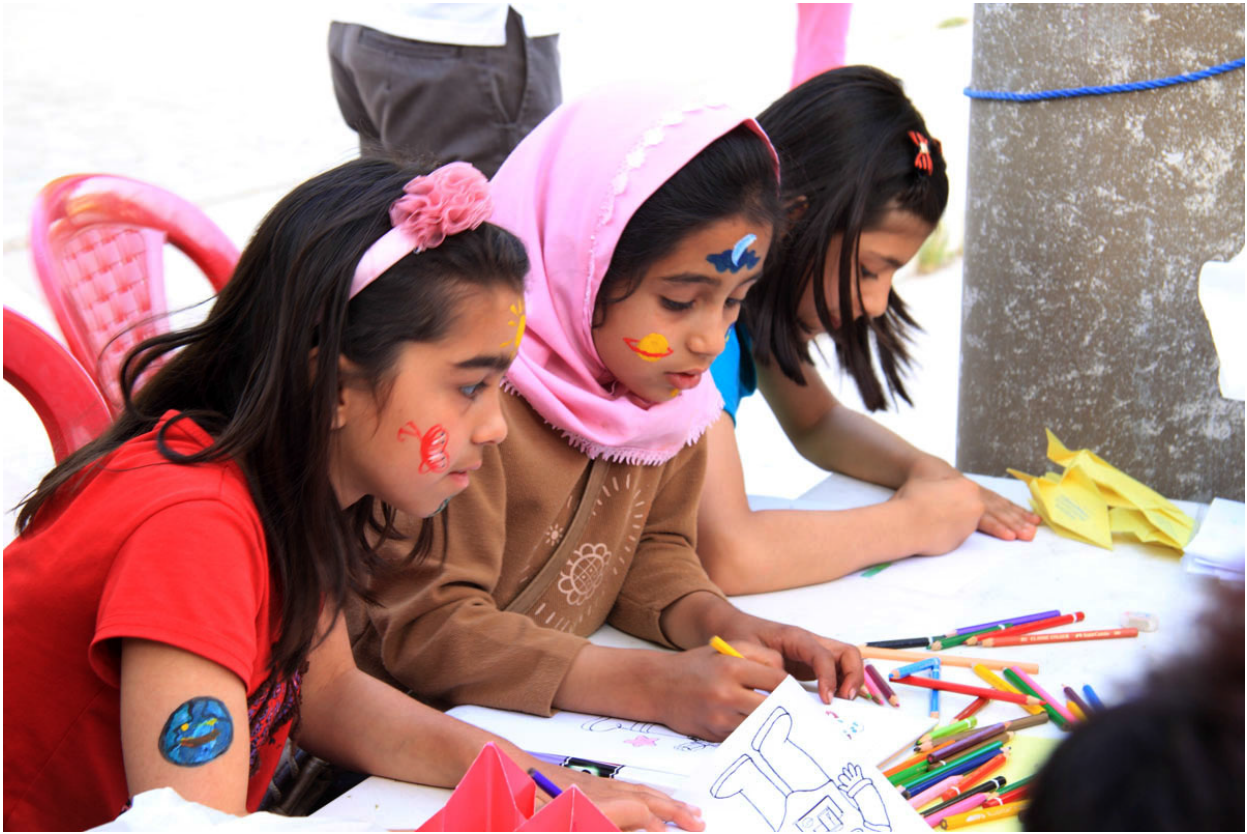

*Figure 5:* Girls taking part in UNAWE Activities in Iran.

**4.2 Case Study 2: Educational Resources - Space Scoop News Service**

An innovative resource developed in the course of implementing the EU Universe Awareness project was 'Space Scoop,' a news service for older children (Russo et al. 2013 - http://www.unawe.org/resources/spacescoop/).

Science is often perceived by children as a dull subject (Allodi, 2002) and an aim of Space Scoop is to change this. By sharing exciting topical astronomical discoveries with them, Space Scoop attempts to inspire children and cultivate their interest in science and technology. Space Scoops are short news articles about astronomical discoveries written in a child-friendly language and accompanied by a stunning astronomical image. Space Scoop is a tool that can be used in many different settings to teach, share, and discuss the latest astronomy news.

The main educational goals of Space Scoop are the following ones:
- Share the excitement of new discoveries with children.
- Enhance understanding of the world and encourage critical thinking.

- Broaden children's minds and cultivate world citizens.

Each Space Scoop begins with a new discovery or observation in the field of astronomy and space science. Press releases of partner organizations (including ESO[1], NAOJ[2], ESA[3], NASA[4]) are translated into a child-friendly language. Similarly, Space Scoop undergoes an internal educational review to assure the highest quality. The final texts are then sent to our group of volunteer translators from all over the globe, where they are translated into up to 33 languages. The final Space Scoops are released to the public at the same time as the original scientific press release.

Space Scoop is a versatile resource which can be adapted to many different formats and digested by many audiences, including children aged 8 and above, the visually-impaired, teachers and science museums. Its wide reach demonstrates that it meets a considerable demand. Among outlets for Space Scoop are several magazines for children, including *Anorak* and *Timbuktu*; the American Association for the Advancement of Science; *Eurekalert!*; a global news service, Goa's *Nahvind Times*; the Dutch science magazine, *Universum*; *Wired.com*; the popular Indonesian website, *Langitselatan*; *National Geographic Indonesia*; and the Slovenian magazine and website, *Portal v Vesolje*. A Space Scoop article, "The Universe is Big, Beautiful...and Mostly Invisible," was featured in an official South African school textbook,[5] which is used in primary schools all over the country. Space Scoop content is also used in the Cambridge University Press English Language Book, *IGCSE English as a Second Language*[6].

Digital media is presently accessible in most parts of the world (Rosen 2010), and a Space Scoop app has been created for Android mobile devices to help widen the audience base. Space Scoop can also be used as the basis for several educational activities. Space Scoop Storytelling uses the articles as the basis for a creative cross-curricula activity that involves astronomy, literature, and presentation skills.

Space Scoop is available in English, Arabic, Dutch, Italian, German, Spanish, Polish, Albanian, Chinese, French, Icelandic, Indonesian, Japanese, Maltese, Norwegian, Portuguese, Romanian, Russian, Sinhalese, Slovenian, Swahili, Turkish, Ukrainian, Vietnamese and Tétum. During 2011 and 2013 there have been 200 Space Scoops released and more than 2100 separate translated Space Scoops.

---

[1] ESO -European Southern Observatory
[2] NAOJ – National Astronomical Observatory of Japan
[3] ESA- European Space Agency
[4] NASA – National Aeronautics and Space Administration
[5] Natural Sciences Grade 8 Textbook, ISBN 0796213720 2013, Heinemann Educational Publishers (South Africa)
[6] http://www.amazon.com/Cambridge-IGCSE-English-Second-Language/dp/1444191624

**Table 1. Space Scoop Impact Numbers from March 2011 to November 2013.**

| Number of Space Scoops published | 196 |
|---|---|
| Languages in which Space Scoops are available | 22 |
| Number of Space Scoops translated | 2108 |
| Number of Space Scoop Partners | 10 |
| Page views on UNAWE website | 100,000 (80% unique page views) |

**5. FUTURE OF UNAWE**

**5.1. Innovation and research potential**

The long-term ambition of the UNAWE team is to expand the reach of UNAWE substantially and stimulate the implementation of UNAWE activities in many more countries, particularly in Africa. To this end there is a need to develop innovative ways for enlarging the reach of the program and new innovative scalable modes for delivery.

One way of expanding the reach of the program is to make courses and demonstration videos for teachers (mini-MOOCs (Massive Open On-line Courses) matched to the language and culture of the targeted country. Another interesting tool for reaching out to children in remote areas is the mobile 'Astrobus' pioneered by UNAWE-Tunisia, an initiative of La Cité des Sciences in Tunis. The bus transports a small telescope, a mini-planetarium and an exhibition throughout Tunisia. Several mobile teacher training pilot projects have been carried out in South America, Africa and India by the *Galileo Mobile* team[7]. The concept of an Astrobus could be applied in many more countries to reach out and deliver UNAWE to remote villages.

There is an additional aspect of UNAWE that is necessary for the future, namely research. A direct research goal of several of UNAWE projects (such as EU-UNAWE) is to evaluate the effectiveness in reaching the UNAWE goals. Evaluation of short-term impact on aspects such as comprehension is relatively simple. However, the long-term effect of such programs in influencing the development of values and attitudes is extremely difficult to evaluate. There are still very few studies that attempt to investigate the long-term effect of education of children on their performance as adults in society. A well-known example is the High/Scope Perry Preschool study (Schweinhart, *et al.*, 2005) that follows the history of two groups of American children from an inner city community through age 40, one of whom has received preschool education. This demonstrates that preschool education has a dramatically positive effect on success. Additional long-term studies in various environments are needed. The global reach of UNAWE, the unique nature of astronomy as a motivational tool, and the educational and social goals of UNAWE make it an excellent test bed for continuing long-term latitudinal studies to investigating the lasting effectiveness of programs that foster the motivation of young children. In general, UNAWE can serve as a unique test bed for interdisciplinary research programs in multiple domains. These include projects to investigate inter-cultural aspects of early child development and the economic effectiveness of motivational educational programs directed at young children in a range of environments and cultures.

---
[7] http://galileo-mobile.org/node/35

**5.2 Sustainability of UNAWE**

Future ambitions of UNAWE are of course contingent on the availability of sufficient funds. On the long term there are several possibilities for obtaining financial support.

1. UNAWE is an integral part of the International Astronomical Union Strategic Plan 'Astronomy for Development' 2010 - 2020, whose implementation is being coordinated by the IAU Office of Astronomy for Development (OAD) in Cape Town. The OAD was inaugurated in 2011 as a joint partnership between the IAU and the South African National Research Foundation. The OAD, with strong backing from the South African authorities, is now well positioned to raise substantial funds. During the next 3 years the OAD will conduct an intensive and widespread fund raising campaign targeted at e.g. international and regional development agencies, multinational companies, philanthropic foundations and individual philanthropists. Support for school education and UNAWE will be important aspects of this campaign.

2. The power of astronomy, in particular, radio astronomy and the associated sophisticated technologies, as a tool for capacity building in Africa has been acknowledged by official resolutions of both the European Parliament and the African Union. It has been decided to site half of the Square Kilometer Array, a next-generation multi-billion dollar global radio telescope in Africa, where it will span 8 African countries. Furthermore the Chinese President Xi Jinping has praised the importance of astronomy as a tool for development. UNAWE is in an excellent position to exploit such developments because it can help inspire the next generation of scientists, engineers and the public with astronomy and thereby help build human capacity globally.

3. UNAWE has substantial support within the European Parliament and the European Commission.

4. Because of its inspirational goals, global reach and demonstrated success, UNAWE is an attractive high-profile program for support by philanthropic foundations and individuals.

**7. UNAWE AND THE "SCIENTIFIC MIND".**

As mentioned at the outset, the philosophy of UNAWE fits well with the precepts of the *Building the Scientific Mind* (BSM) approach to education. However, it may be questioned whether the term "scientific mind" is a misnomer and could be counterproductive in achieving the BtSM goal of reforming the education system. Although derived from the Latin *scientia* or 'knowledge,' in modern English science is usually taken as referring to disciplines that seek to explain the phenomena of the material universe and in this it is distinguished from the humanities.

In 1959, C.P. Snow delivered a famous set of lectures that drew attention to the two cultures (Snow & Collini, 2012). He spoke of scientists who could scarcely struggle through a novel by Dickens, but more importantly of humanities professors who were ignorant of the Second Law of Thermodynamics, who sneered at science as an inferior branch of learning that no really cultured person needed to trouble with. "If the scientists have the future in their bones," he claimed, "then the traditional culture responds by wishing the future did not exist" (p. 11). "There is only one way out of all this," Snow argued. "It is, of course, by rethinking our education" (p. 18).

BSM rightly advocates the mobilization of children's curiosity from the earliest age on so as to stimulate their ability to think rationally and to develop their sense of ethical values. Surely, cultivating BtSM values and skills is equally important for the appreciation and pursuit of art, music and literature and for the correct interpretation of history as for doing science? BtSM is a holistic educational concept designed to stimulate the human mind as a whole and stimulate rationality and creativity and not just the ability to do science.

In recent years the need for radical changes in approaches to education has received a new impetus from several directions. Search engines and the web have resulted in a revolution in our access to knowledge, so that teaching facts is regarded as not nearly as important as was previously the case (Wagner, 2008). The skill sets that are needed to function optimally in the modern global economy and realize a person's full potential are creative ones which stimulate the recognition of links between seemingly unrelated facts and help one to build innovatively from existing facts. Above all, motivating children to learn is crucial and it can be argued that the development of an ethical value system is critical to the survival of human civilization. We have attempted to demonstrate in this chapter that the use of astronomy, space and the philosophy of Universe Awareness can contribute substantially to the needed transformation of education globally and to the BtSM approach to education.

## 8. REFERENCES.